\documentclass[reprint,aps,pra,superscriptaddress,showpacs,footinbib,longbibliography,amsmath]{revtex4-1}
\pdfoutput=1
\usepackage{graphicx}
\usepackage{dcolumn}
\usepackage{bm}

\usepackage{color}
\usepackage[colorlinks,bookmarks=false,citecolor=darkblue,linkcolor=red,urlcolor=blue]{hyperref}

\definecolor{darkred}{rgb}{0.7,0.0,0.0}

\definecolor{darkblue}{rgb}{0,0.02,0.45}

\definecolor{darkgreen}{rgb}{0.02,0.45,0.0}

\definecolor{violet}{rgb}{0.8,0.2,0.6}

\graphicspath{{../}}

\begin{document}
\title{Persistent spin dynamics in the pressurized spin-liquid candidate YbMgGaO$_4$}

\author{Mayukh Majumder}
\affiliation{Experimental Physics VI, Center for Electronic Correlations and Magnetism, University of Augsburg, 86159 Augsburg, Germany}

\author{Gediminas Simutis}
\affiliation{Laboratory for Muon Spin Spectroscopy, Paul Scherrer Institut, 5232 Villigen PSI, Switzerland}

\author{Ines E. Collings}
\affiliation{European Synchrotron Radiation Facility, 71 Avenue des Martyrs, 38000 Grenoble, France}

\author{Jean-Christophe Orain}
\affiliation{Laboratory for Muon Spin Spectroscopy, Paul Scherrer Institut, 5232 Villigen PSI, Switzerland}

\author{Tusharkanti Dey}
\author{Yuesheng Li}
\author{Philipp Gegenwart}
\author{Alexander A. Tsirlin}
\email{altsirlin@gmail.com}
\affiliation{Experimental Physics VI, Center for Electronic Correlations and Magnetism, University of Augsburg, 86159 Augsburg, Germany}


\begin{abstract}
Single-crystal x-ray diffraction, density-functional band-structure calculations, and muon spin relaxation ($\mu$SR) are used to probe pressure evolution of the triangular spin-liquid candidate YbMgGaO$_4$. The rhombohedral crystal structure is retained up to at least 10\,GPa and shows a nearly uniform compression along both in-plane and out-of-plane directions, whereas local distortions caused by the random distribution of Mg$^{2+}$ and Ga$^{3+}$ remain mostly unchanged. The $\mu$SR data confirm persistent spin dynamics up to 2.6\,GPa and down to 250\,mK with no change in the muon relaxation rate. Longitudinal-field $\mu$SR reveals power-law behavior of the spin-spin autocorrelation function, both at ambient pressure and upon compression. 
\end{abstract}

\maketitle

\section{Introduction}
Spin-liquid states in frustrated magnets are nowadays actively studied as hosts for unconventional excitations representing magnetic monopoles~\cite{balents2010,gingras2014} and other exotic quasiparticles~\cite{savary2017,hermanns2018}. One relatively unexplored aspect in this field is the evolution of spin-liquid materials under pressure and the change in spin dynamics caused by tuning magnetic interactions via lattice compression. 

Here, we focus on the spin-liquid candidate YbMgGaO$_4$~\cite{li2015a,li2015b,li2020} that recently evolved as a unique triangular antiferromagnet with the robust three-fold symmetry, persistent spin dynamics, and a broad continuum of (potentially fractionalized) magnetic excitations. The crystal structure of this compound features triangular layers of the pseudospin-$\frac12$ Yb$^{3+}$ ions that are well separated by slabs of non-magnetic Mg$^{2+}$ and Ga$^{3+}$ (Fig.~\ref{fig:local}a). Thermodynamic measurements~\cite{li2015a,li2015b} and muon spin relaxation ($\mu$SR)~\cite{li2016} suggest the absence of magnetic order down to at least 50\,mK. Weak spin freezing around 100\,mK was indicated by ac-susceptibility data~\cite{ma2018}, although it involves only a tiny amount of the magnetic entropy~\cite{li2015a} and leaves no signatures in either dc-susceptibility~\cite{li2019} or $\mu$SR~\cite{li2016}. 

Magnetic excitations of YbMgGaO$_4$ form a broad continuum that can be interpreted in terms of gapless spinons~\cite{shen2016,chen2017a,chen2017b,shen2018} or as arising from short-range valence bonds~\cite{li2017b,li2019}, the latter suggestion being particularly interesting, as it makes a direct link to Anderson's pioneering work at the outset of the spin-liquid research~\cite{anderson1973}. Continuum features were also reported for other Yb$^{3+}$ triangular antiferromagnets~\cite{ding2019,ma2020,dai2020}, but they show a different distribution of the spectral weight and probably have a separate origin. Indeed, YbMgGaO$_4$ is known to be strongly affected by structural randomness that arises from the random distribution of Mg$^{2+}$ and Ga$^{3+}$ between the Yb$^{3+}$ layers and modulates magnetic interactions via random crystal electric fields acting on Yb$^{3+}$~\cite{li2017a,paddison2017}. No such randomness occurs in other Yb-based triangular antiferromagnets.


In the following, we explore the effect of hydrostatic pressure on the structure and magnetism of YbMgGaO$_4$, and juxtapose the behavior of this material with the evolution of other spin-liquid candidates upon compression. We show that in YbMgGaO$_4$ spins remain dynamic up to at least 2.6\,GPa, and quantify associated structural changes for an eventual comparison across different classes of spin-liquid materials.


\section{Crystal structure}
\subsection{Average structure}
Room-temperature x-ray diffraction (XRD) data were collected at the ID15B beamline of the European Synchrotron Radiation Facility (ESRF) between ambient pressure and 10\,GPa. A diamond anvil cell loaded with He gas and a small single crystal of YbMgGaO$_4$ from the batch reported in Ref.~\onlinecite{li2015b} were used for the experiment. 

Three positional parameters ($z$-coordinates) and three independent atomic displacement parameters were refined for Mg/Ga, O1, and O2. The position of Yb was fixed at $(0,0,0)$, with its thermal ellipsoid refined anisotropically to account for local displacements caused by the random distribution of Mg and Ga. Altogether, 9 structural parameters were refined from about 120 symmetry-independent reflections collected at each pressure point~\cite{suppl}. Details of the data collection and structure refinement are given in Appendix~\ref{app:xrd}.

\begin{figure}
\includegraphics{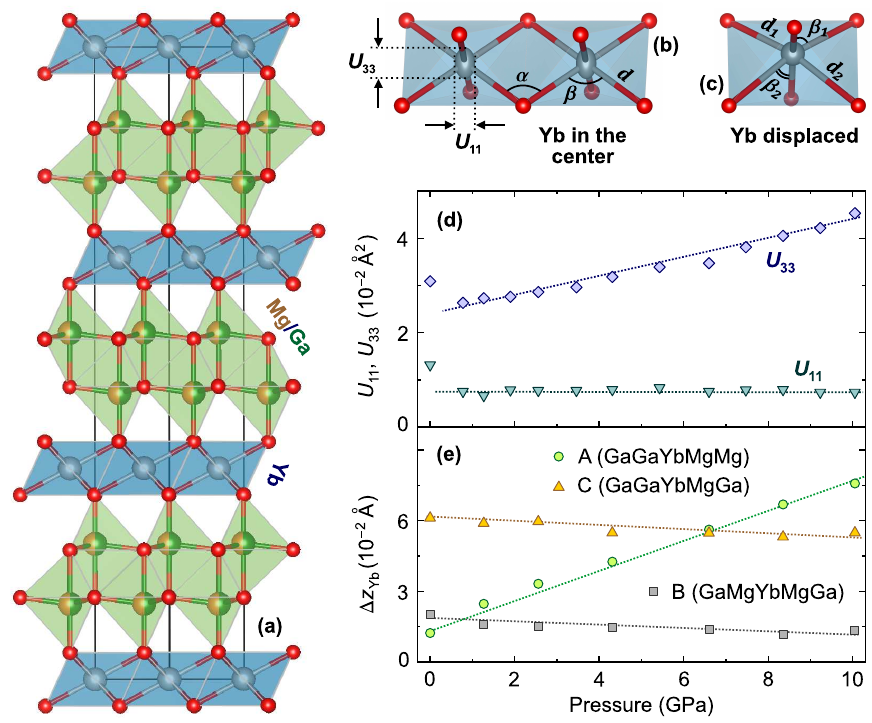}
\caption{\label{fig:local}
(a) Crystal structure of YbMgGaO$_4$ with Mg$^{2+}$ and Ga$^{3+}$ randomly occupying the position in the trigonal bipyramids; (b) local structural parameters, the Yb--O distance ($d$) and Yb--O--Yb ($\alpha$) or O--Yb--O ($\beta$) angles with $\alpha=\beta$; (c) distortion of the YbO$_6$ octahedron caused by an asymmetric distribution of the Mg and Ga atoms; (d) experimental atomic displacement parameters of Yb; (e) displacements of the Yb atoms obtained for different local configurations shown in Fig.~\ref{fig:deformation}.
}
\end{figure}

The $R\bar 3m$ symmetry of YbMgGaO$_4$ keeps all Yb--O distances equal but allows a trigonal distortion of the YbO$_6$ octahedra with the O--Yb--O angle $\beta$ deviating from $90^{\circ}$. Incidentally, this angle is equal by symmetry to the Yb--O--Yb bridging angle $\alpha$, which is responsible for superexchange interaction (Fig.~\ref{fig:local}b). Two structural parameters, one distance and one angle, are thus sufficient to characterize both the local geometry of Yb$^{3+}$ and the geometry of the nearest-neighbor exchange pathway within the average structure.

\begin{figure}
\includegraphics[scale=0.95]{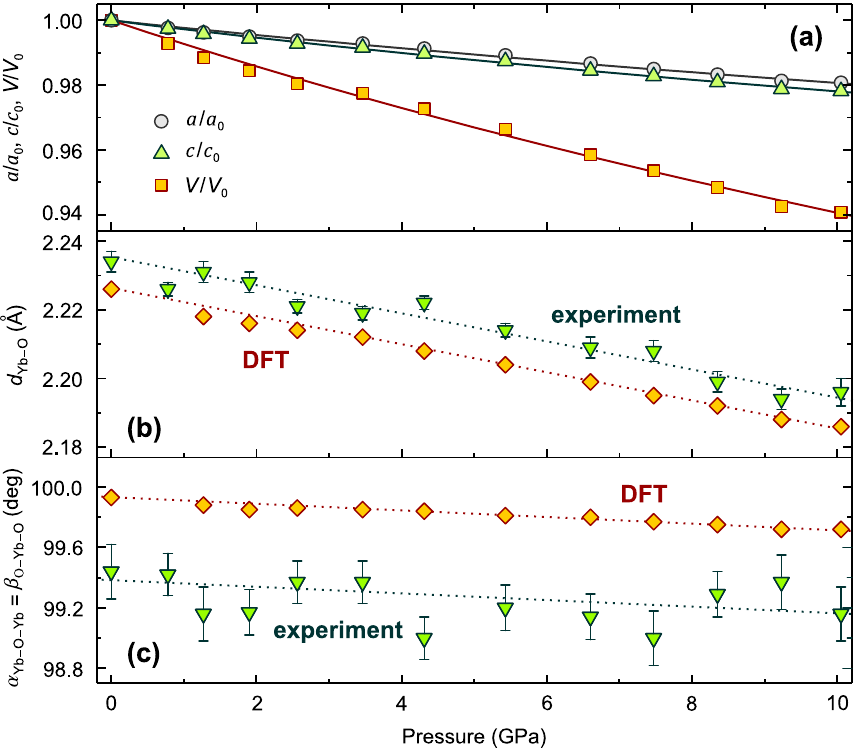}
\caption{\label{fig:structure}
Pressure-induced changes of the average structure determined using single-crystal XRD: (a) lattice parameters and unit cell volume; (b) Yb--O distance $d$; (c) Yb--O--Yb / O--Yb--O angles ($\alpha=\beta$). The lines in panel (a) are fits with the second-order Murnaghan equation of state. Panels (b) and (c) additionally show the geometrical parameters obtained from DFT structure relaxations detailed in Appendix~\ref{app:dft}, and dotted lines are guide-for-the-eye.
}
\end{figure}

No changes in the crystal symmetry were observed in our experiment. Pressure evolution of the $a$ and $c$ lattice parameters demonstrates similar compressibility along the different crystallographic directions (Fig.~\ref{fig:structure}a). The fit with the second-order Birch-Murnaghan equation of state,
\begin{equation}
 p(V)=\frac{3B_0}{2}\left[\left(\frac{V_0}{V}\right)^{\frac73}-\left(\frac{V_0}{V}\right)^{\frac53}\right],
\end{equation}
using the \texttt{EosFit} routine~\cite{eosfit} returned the bulk modulus of $B_0=142(2)$\,GPa and the unit-cell volume of $V_0=253.12(7)$\,\r A$^3$. A similar fit for individual lattice parameters using $a^3$ or $c^3$ in place of $V$ yields $B_{0,a}=151(3)$\,GPa and $B_{0,c}=130(2)$\,GPa suggesting a nearly isotropic compression of the structure. YbMgGaO$_4$ is more compressible than Yb$_2$Ti$_2$O$_7$ ($B_0=219$\,GPa~\cite{mishra2012}) and other rare-earth pyrochlores that typically feature the bulk moduli in excess of 200\,GPa~\cite{turner2017}. 

The Yb--O distances shrink by about 0.6\% at 2.6\,GPa (the highest pressure of our $\mu$SR experiment, Sec.~\ref{sec:musr}) and by 1.7\% at 10\,GPa (Fig.~\ref{fig:structure}b). The $\alpha=\beta$ angle shows a weak downward trend only, with a larger error bar caused by the lower accuracy in the determination of oxygen position due to the low scattering power of oxygen. In order to confirm this downward trend, we relaxed the experimental structures using density-functional (DFT) band-structure calculations. As DFT can not treat the Mg/Ga disorder explicitly, ordered structural models have to be used, as explained in Appendix~\ref{app:dft}. This approximation leads to a constant offset between the DFT results and experiment. Nevertheless, not only the qualitative trends but also the slope are well reproduced (Fig.~\ref{fig:structure}). We thus confirm that the Yb--O--Yb angle decreases under pressure. Compared to the ambient-pressure value, it changes by $0.07^{\circ}$ at 2.6\,GPa and by $0.2^{\circ}$ at 10\,GPa.

\subsection{Local structure} 
Moderate changes of the average crystal structure are accompanied by a strong elongation of the Yb thermal ellipsoid. The in-plane displacements characterized by $U_{11}$ are not affected by pressure, whereas the out-of-plane displacement component increases by 70\% (Fig.~\ref{fig:local}d). This out-of-plane displacement has been previously linked to the local distortions of the YbO$_6$ octahedra caused by the random (and, generally, asymmetric) distribution of the differently charged Mg$^{2+}$ and Ga$^{3+}$ ions around Yb$^{3+}$~\cite{li2017a}. At first glance, the increase in $U_{33}$ implies a strong enhancement of the structural randomness under pressure, but the actual situation is more complex.

\begin{figure*}
\includegraphics{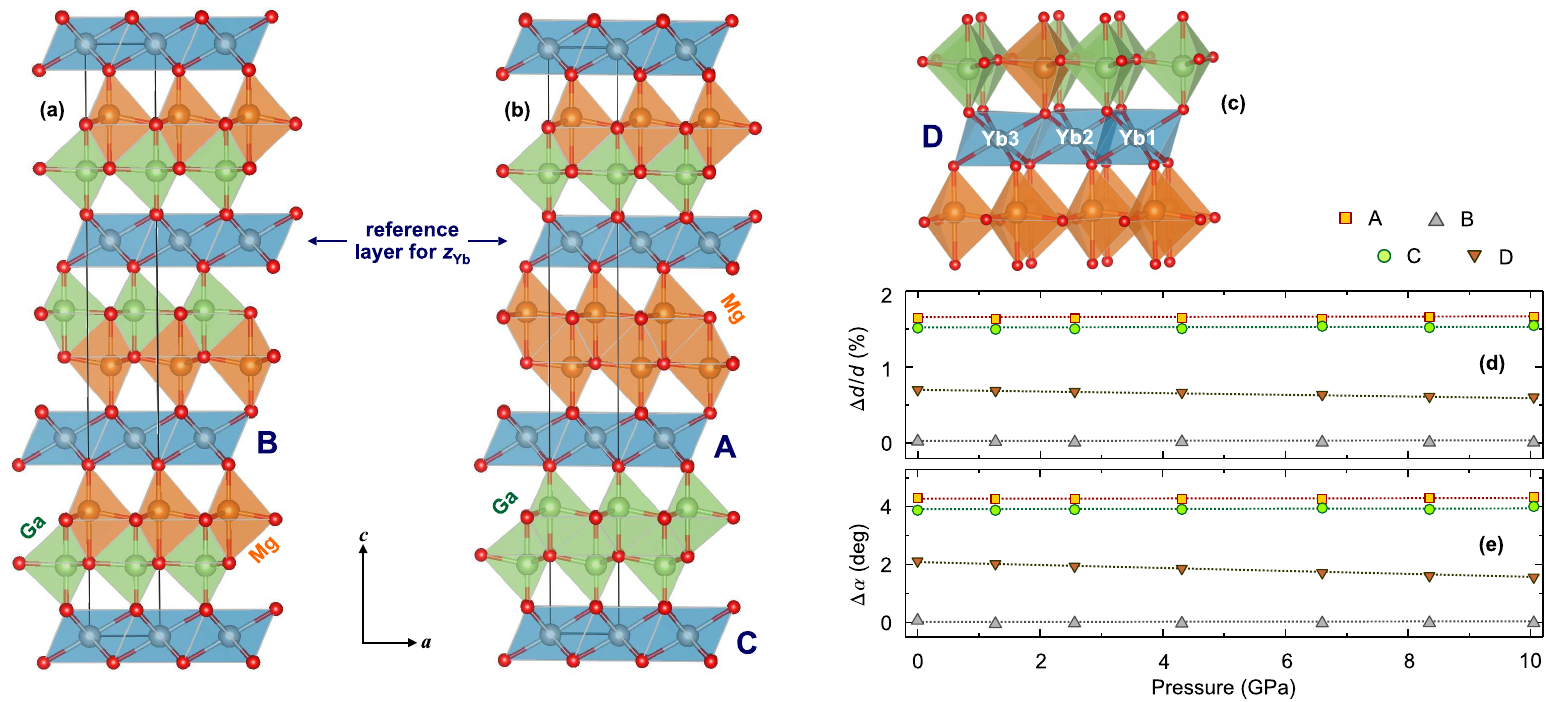}
\caption{\label{fig:deformation}
Ordered models for evaluating pressure evolution of the local structure: (a) and (b) layer-by-layer ordering of Mg$^{2+}$ and Ga$^{3+}$ leading to the Yb layers A, B, and C with same YbO$_6$ octahedra throughout each layer; (c) a more complex model with layer D containing non-equivalent YbO$_6$ octahedra; (d) relative deformations of the octahedra obtained as $(d_1-d_2)/(d_1+d_2)$ for layers A, B, and C and as a difference between the average $d$ for the Yb1 and Yb2 octahedra in layer D; (e) angular deformation obtained as $\Delta\alpha=\Delta\beta=\beta_2-\beta_1$ for layers A, B, and C and as the difference in the Yb1--Yb2 and Yb2--Yb3 superexchange angles for layer D.
}
\end{figure*}

We analyze pressure evolution of the local structure by DFT relaxations for several ordered models of YbMgGaO$_4$. First, we consider the models of Ref.~\onlinecite{li2017a} that were previously used to interpret crystal-field excitations of Yb$^{3+}$, and evaluate $\Delta z_{\rm Yb}$ as the displacements of the Yb atoms relative to each other. These displacements indicate the spread of the Yb electron density, which is gauged by the $U_{33}$ parameter of the average structure. 

Three scenarios visualized in Fig.~\ref{fig:deformation}a,b are considered, with the Yb$^{3+}$ ions sandwiched between: i) two Ga$^{3+}$ ions on one side and two Mg$^{2+}$ ions on the other side (octahedra A); ii) one Ga$^{3+}$ and one Mg$^{2+}$ ion on each side (octahedra B); iii) two Ga$^{3+}$ ions on one side and a combination of Ga$^{3+}$ and Mg$^{2+}$ on the other side (octahedra C). Pressure has a strong effect on the Yb position in the octahedra A and nearly no effect on the Yb atoms in B and C (Fig.~\ref{fig:local}e). Surprisingly, this effect is a shift of the whole octahedron A along the $c$ direction without any change in the octahedron itself. Using the geometrical parameters shown in Fig.~\ref{fig:local}c, we determine that for the octahedra A the deformation expressed by $(d_2-d_1)$ changes from 1.65\,\% at 0\,GPa to 1.66\,\% at 10\,GPa, whereas $(\beta_2-\beta_1)$ evolves from $4.29^{\circ}$ to $4.32^{\circ}$ (Fig.~\ref{fig:deformation}d,e). The changes in the octahedra B and C are equally small. Since $\alpha=\beta$ in our material, the distribution of the \mbox{Yb--O--Yb} angles also does not change with pressure. Therefore, neither local randomness at each Yb site nor randomness of the exchange couplings should be affected by pressure.

We also consider a different scenario where two adjacent Yb atoms within the same layer have a different local environment. In layer D (Fig.~\ref{fig:deformation}c), Yb1 is surrounded by 3 Ga atoms on one side and 3 Mg atoms on the other side. In contrast, Yb2 and Yb3 are surrounded by 2 Ga and 4 Mg atoms each. These dissimilar local environments lead to different $\alpha_1$ and $\alpha_2$ for the Yb1--Yb2 and Yb2--Yb3 superexchange pathways and reflect the effect of structural randomness on nearest-neighbor magnetic interactions in YbMgGaO$_4$. However, the difference between these two pathways appears to be only weakly pressure-dependent and decreases upon compression (Fig.~\ref{fig:deformation}e).

Our results suggest that the visible enhancement of $U_{33}$ under pressure is \textit{not} related to the increased distortions around the Yb$^{3+}$ ions. It rather indicates a change in the overall position of the octahedra A that may be explained by the accumulation of different charges above (Ga$^{3+}$) and below (Mg$^{2+}$) the Yb layer. But the key result at this juncture is that such a change reflects a re-arrangement within the Mg/Ga slabs and does not affect the structure of the magnetic layer itself. The YbO$_6$ octahedra undergo a uniform compression and simply keep the deformation that they had at ambient pressure. The randomness effect on the exchange couplings should be largely unchanged or even decrease under pressure.

\section{Spin dynamics} 
\label{sec:musr}
Pressure evolution of spin dynamics is probed by muon spin relaxation ($\mu$SR). The experiments were performed at the GPD and Dolly spectrometers at the Paul Scherrer Institute (PSI) at ambient pressure and at the GPD spectrometer under hydrostatic pressure~\cite{khasanov2016} down to 250\,mK on a polycrystalline sample of YbMgGaO$_4$. The data were collected in zero field (ZF) and in the longitudinal-field (LF) mode, where the magnetic field was applied parallel to the spin of the implanted muons. 

\subsection{ZF $\mu$SR}
\begin{figure}
\includegraphics[width=8cm]{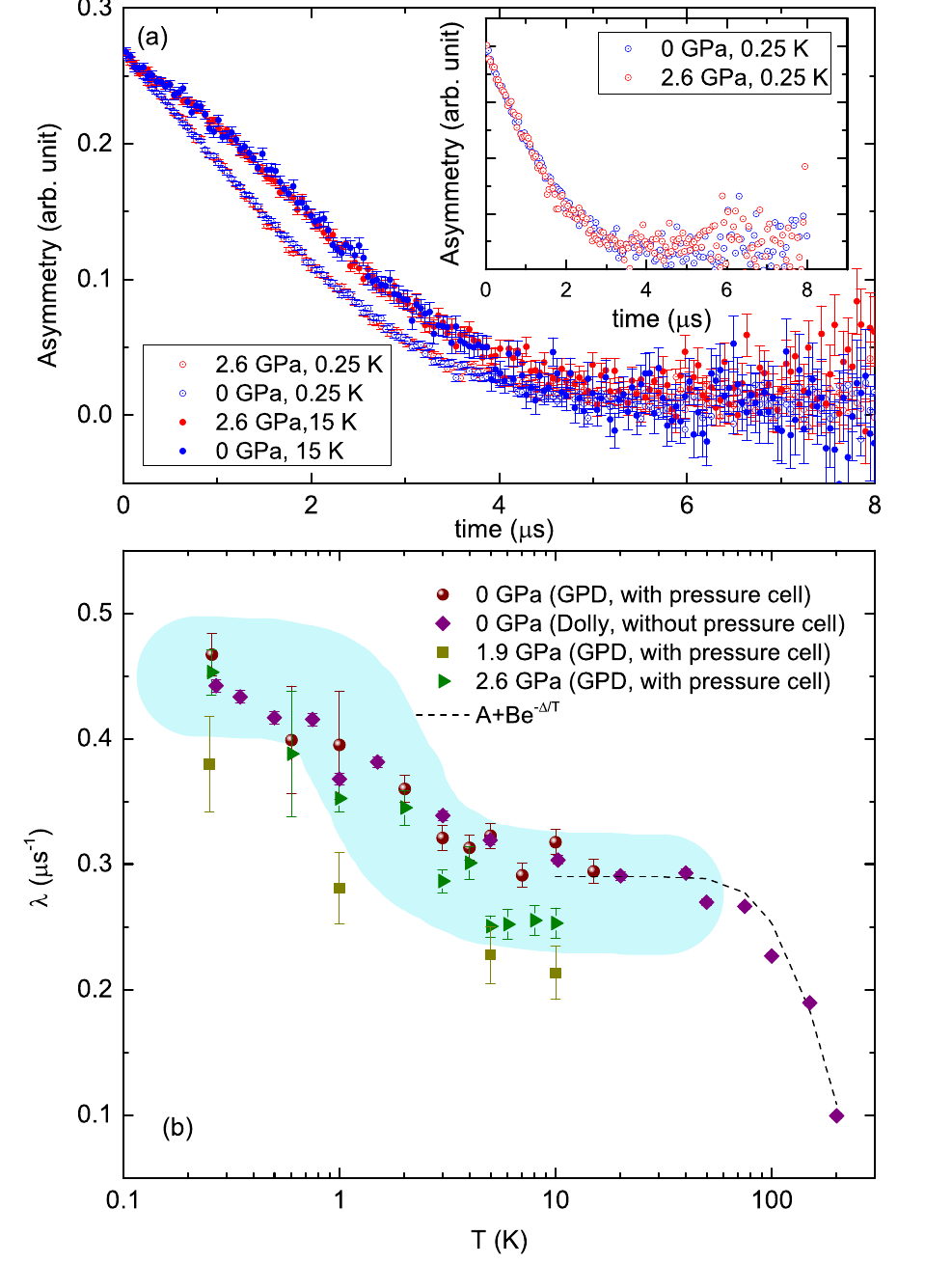}
\caption{\label{fig:asymmetry}
(a): Zero-field $\mu$SR spectra measured at 0 and 2.6\,GPa at both 15\,K and 250\,mK. The inset shows the sample contribution after subtracting $f_{\rm PC}A_{\rm PC}$ of Eq.~\eqref{eq:asym}. (b): Temperature dependence of the zero-field muon relaxation rate measured at different pressures. The dashed line is the fit with the activated behavior above 40\,K. 
}
\end{figure}

ZF $\mu$SR time spectra measured at ambient pressure and at 2.6\,GPa, the highest pressure of our experiment, are compared in Figure~\ref{fig:asymmetry}. The similar behavior of the $\mu$SR time spectra indicates no change of the magnetic ground state, and the absence of oscillations excludes pressure-induced magnetic ordering in YbMgGaO$_4$ up to at least 2.6\,GPa. To estimate the temperature dependence of the relaxation rate $\lambda$, we fitted the ZF-$\mu$SR time spectra by 
\begin{equation}
A(t) = f_{\rm PC}A_{\rm PC} + (1-f_{\rm PC})e^{-\lambda t}
\label{eq:asym}\end{equation}
where $A_{\rm PC}$ is the pressure-cell contribution described in Appendix~\ref{app:musr}. The fraction of signal coming from the pressure cell was $f_{\rm PC}=0.7$ at 0\,GPa and 2.6\,GPa, and $f_{\rm PC}=0.5$ at 1.9\,GPa, where a different cell was used in an attempt to reduce the background.

The zero-field muon relaxation rate $\lambda$ is temperature-independent between 4 and 40\,K (Fig.~\ref{fig:asymmetry}). Its increase below 4\,K indicates the onset of spin-spin correlations that fully develop around 0.8\,K, where $\lambda$ flattens out and remains temperature-independent upon further cooling. This temperature evolution is essentially similar to the ambient-pressure $\mu$SR data reported in Ref.~\onlinecite{li2016} and remains unchanged at 2.6\,GPa (Fig.~\ref{fig:asymmetry}). Even absolute values of $\lambda$ are the same as at ambient pressure within the error bar.

Above 40\,K, $\lambda$ shows a steady decreasing trend described by an activated behavior $\lambda = A+B\,e^{-\Delta/T}$ with $\Delta = 320\pm20$\,K, which is reminiscent of the lowest crystal-field excitation energy of about 450\,K~\cite{li2017a}. This observation suggests that at high temperatures the relaxation is governed by an Orbach process~\cite{orbach1961} involving the excited crystal-field doublets of Yb$^{3+}$.

\subsection{LF $\mu$SR}
LF measurements complement the data obtained in zero field. Even a longitudinal field, which is more than 10 times higher than the local static field estimated from the low-temperature value of $\lambda$, does not decouple the muon relaxation, suggesting that the spins are dynamic in nature as has been seen at ambient pressure~\cite{li2016}. 

\begin{figure}
\includegraphics[width=8cm]{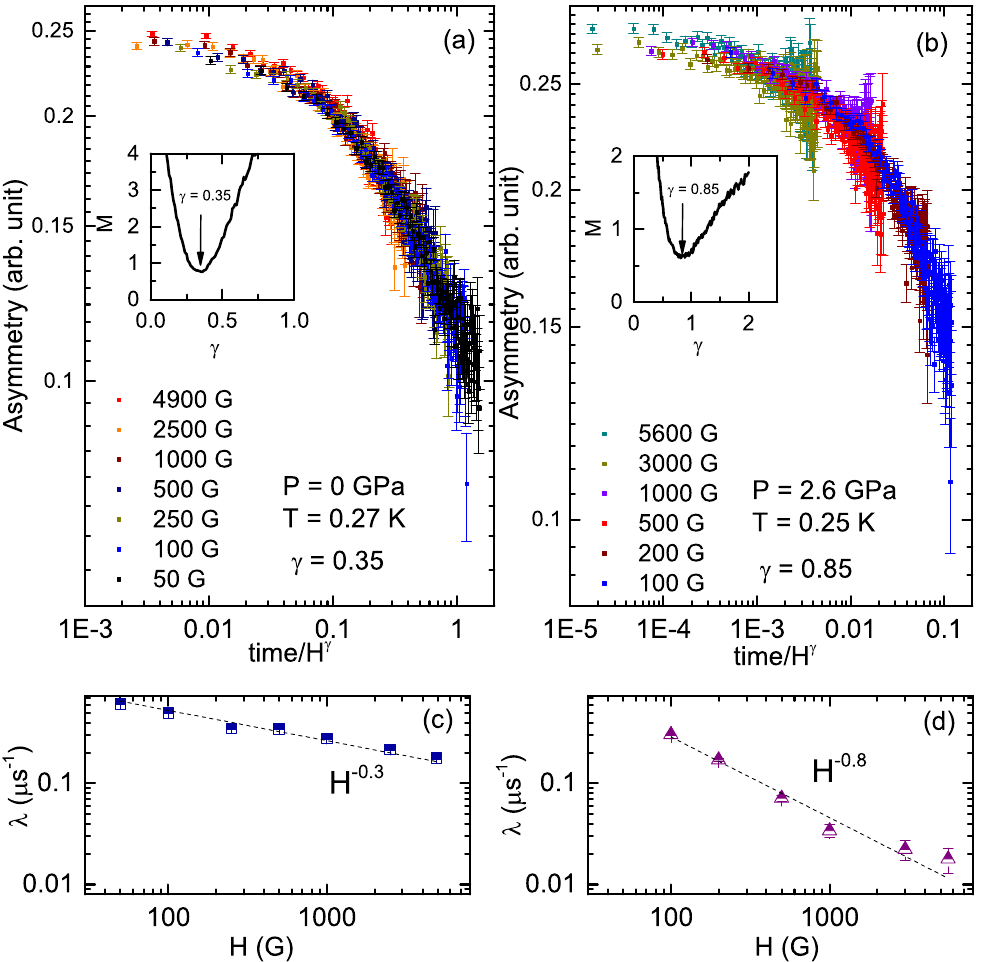}
\caption{\label{fig:scaling}
(a,b) Universal scaling of the LF-$\mu$SR data measured at 0\,GPa and 270\,mK (left, $\gamma=0.35$) and 2.6\,GPa and 250\,mK (right, $\gamma=0.85$), The insets show the data mismatch function as a function of $\gamma$. (c,d) Power-law behavior of the muon relaxation rate, $\lambda(H)\sim H^{-\gamma}$.
}
\end{figure}

The LF data measured in different fields follow universal scaling initially proposed by Keren~\textit{et al.}~\cite{keren1996} for systems with glassy dynamics. This scaling manifests itself in the power-law behavior of the muon relaxation rate $\lambda(H)$ extracted by fitting individual LF curves with a stretched exponential $e^{-[\lambda(H)t]^{\beta}}$ supplied with the time-independent and field-independent background, which originates from the sample holder and pressure cell. The ensuing values of $\lambda$ follow $\lambda(H)\sim H^{-\gamma}$ with $\gamma\simeq 0.3$ at 0\,GPa and 0.8 at 2.6\,GPa (Fig.~\ref{fig:scaling}). 

This behavior is confirmed by the scaling of muon asymmetry plotted against $t/H^{\gamma}$. To this end, the data points from all fields up to $t=6$\,$\mu$s~\footnote{The data above $t=6$\,$\mu$s were excluded due to their large error bars.} are arranged with increasing $t/H^\gamma$ for every value of $\gamma$. An empirical data mismatch function is calculated by taking the difference between the neighboring points and weighing them by the corresponding error bars. This mismatch function is defined as
\begin{equation}
M = \dfrac{1}{N}\sum_i^N \dfrac{(A_i-A_{i+1})^2}{(\delta_i-\delta_{i+1})^2}
\end{equation}
where $N$ is the number of data points, and $A_i$ and $\delta_i$ correspond to the asymmetry and error bar of the $i$-th data point, respectively. The lowest value of the mismatch function is obtained at $\gamma=0.35$ (0\,GPa) and $\gamma=0.85$ (2.6\,GPa) that produce the universal scaling over at least three orders of magnitude in $t/H^{\gamma}$ and show excellent agreement with the $\gamma$ values from the analysis of $\lambda(H)$. 

The scaling may not hold in the whole time and field range -- e.g., at very short times comparable to the width of the muon pulse~\cite{keren1996} or in very high fields that affect spin dynamics -- but the scaling over three orders of magnitude (Fig.~\ref{fig:scaling}) is congruent with all earlier observations~\cite{keren2004,maclaughlin2004} and serves as a robust evidence of the power-law behavior $q(t)\sim t^{\gamma-1}$ of the dynamic spin-spin autocorrelation function $q(t) =\langle S_i(t)\cdot S_i(0)\rangle$. It indicates collective dynamics, as for example in spin-glass systems where $\gamma$ increases toward 1.0 upon approaching the freezing temperature from above~\cite{keren1996}. 


\section{Discussion and summary} 
YbMgGaO$_4$ is remarkably insensitive to pressure and thus different from other spin-liquid candidates. For example, herbertsmithite~\cite{mendels2010,norman2016} becomes magnetically ordered at 2.5\,GPa~\cite{kozlenko2012}, whereas stoichiometric Yb$_2$Ti$_2$O$_7$~\cite{rau2019} shows pressure-induced magnetic order already at 0.1\,GPa~\cite{kermarrec2017}. These materials have different compressibilities, with the bulk modulus of $B_0=70$\,GPa in herbertsmithite~\cite{kozlenko2012} and $B_0=219$\,GPa in Yb$_2$Ti$_2$O$_7$~\cite{mishra2012}. YbMgGaO$_4$ shows an intermediate value of 142\,GPa and is certainly comparable to other spin-liquid candidates as far as elastic properties are concerned. Its structural parameters are affected by pressure. The persistence of spin dynamics will then indicate that structural changes influence exchange couplings only weakly, whereas the material lies sufficiently far away from the phase boundary between the spin-liquid and magnetically ordered states. Alternatively, spin dynamics may not be caused by frustration that controls this phase boundary, and originate from structural randomness, which is not affected by pressure. 

Triangular antiferromagnets develop $120^{\circ}$ order or stripe order depending on the extent of exchange anisotropy~\cite{luo2017} and on the ratio $J_2/J_1$ of the second-neighbor to nearest-neighbor exchange coupling~\cite{zhu2018,maksimov2019}. The formation of a spin liquid is mostly controlled by $J_2/J_1$, with the complete suppression of magnetic order between $J_2/J_1\simeq 0.07$ and 0.15~\cite{zhu2018,maksimov2019,li2020}. Experimental estimates of this parameter vary in a broad range between 0.0 and 0.26~\cite{li2020} and would place YbMgGaO$_4$ in any of the regions on the phase diagram: $120^{\circ}$ order~\cite{li2015b}, stripe order~\cite{paddison2017,zhu2017}, or the boundary between the latter and the spin-liquid phase~\cite{zhang2018}. 

External pressure of 2.6\,GPa leads to a shortening of the Yb--O distance $d$ by 0.013\,\r A and a reduction in the Yb--O--Yb angle $\alpha$ by $0.07^{\circ}$ (Fig.~\ref{fig:structure}). Using superexchange theory of Ref.~\onlinecite{rau2018}, we estimate that the change in the angle increases $J_1$ by 6\,\%, whereas the shortening of the Yb--O distance increases $J_1$ by another 5\,\% assuming exponential dependence of Slater-Koster parameters on the interatomic distance. Overall, we expect that $J_1$ increases by 11\,\%, which is similar in magnitude to pressure-induced changes in Cu-based quantum magnets~\cite{zayed2017,zvyagin2019}. No significant changes are expected for $J_2$, because long-range couplings are less sensitive to the structural geometry. In YbMgGaO$_4$, the Yb--O--O angles decrease by as little as $0.11^{\circ}$ upon compression to 2.6\,GPa. In contrast, in a Cu-based quantum magnet even the $2^{\circ}$ change of the respective angle reduces the coupling by 20\,\% only~\cite{prishchenko2017}.

We conclude that $J_1$ increases, $J_2$ is roughly unchanged, and the $J_2/J_1$ ratio of YbMgGaO$_4$ should be reduced under pressure. Assuming the parametrization of Ref.~\onlinecite{zhang2018} with $J_2/J_1=0.18$, it would imply that pressure pushes the material into the spin-liquid region, in agreement with the persistence of spin dynamics observed experimentally. Alternatively, and perhaps even more likely, the lack of significant changes under pressure may indicate that spin dynamics is triggered by structural randomness and thus unaffected by pressure.

The only change we observe is the increase in the $\gamma$ parameter that describes scaling behavior in longitudinal fields. For a system with glassy dynamics the increase in $\gamma$ would reflect an evolution toward a frozen state~\cite{keren1996} that was indeed proposed for YbMgGaO$_4$ below 100\,mK (at ambient pressure) based on the ac-susceptibility data~\cite{ma2018}. However, neither dc-susceptibility~\cite{li2019} nor $\mu$SR~\cite{li2016} support bulk spin freezing at ambient pressure. The scaling behavior may be in fact unrelated to glassy dynamics, and indeed it was also reported for a variety of systems with critical fluctuations influenced by structural disorder~\cite{maclaughlin2004,keren2004}. On the other hand, the very presence of this scaling serves as an additional argument for collective spin dynamics caused by the structural randomness, and the importance of this randomness for the spin-liquid behavior of YbMgGaO$_4$. 

Altogether, hydrostatic pressure leads to a uniform compression of the YbMgGaO$_4$ structure with the reduction in the Yb--O distances and Yb--O--Yb angles, whereas local distortions of the YbO$_6$ octahedra and consequent randomness effects are nearly unchanged. Spin dynamics is not affected by pressure and appears to be collective yet influenced by the structural randomness. This puts YbMgGaO$_4$ into the group of materials, where randomness effects can be integral to the spin-liquid formation, but collective spin dynamics is nevertheless observed. 

\acknowledgments
We acknowledge ESRF for provision of beamtime at ID15B. IEC thanks J. Jacobs for the He gas load. Part of this work is based on experiments performed at the Swiss Muon Source S$\mu$S, Paul Scherrer Institute, Villigen, Switzerland. Swiss National Science Foundation has supported the work of G.S. (Grants No. 200021-175935 and Mobility grant P2EZP2-178604). The work in Augsburg was supported by the German Research Foundation (DFG) via the Project No. 107745057 (TRR80) and by the Federal Ministry for Education and Research through the Sofja Kovalevkaya Award of Alexander von Humboldt Foundation (AAT). 

\appendix
\section{X-ray diffraction}
\label{app:xrd}
Membrane-driven LeToullec-type diamond anvil cells (DACs) were used, equipped with Boehler-Almax anvils. Stainless steel was used as the gasket material, and helium was loaded as the pressure-transmitting medium. Diffraction patterns were collected with a Mar555 flat-panel detector using steps of $0.5^{\circ}$ oscillations over a total $\omega$ scan range of $76^{\circ}$ about the vertical axis. The pressures were measured using the ruby fluorescence method. Lattice parameter determination and integration of the reflection intensities were performed using the \texttt{CrysAlisPro} software~\cite{crysalispro}. Structures were refined using \texttt{ShelxL}~\cite{shelxl} within the \texttt{ShelXle}~\cite{shelxle} graphical interface. 

Details of the data collection and structure refinement at 10\,GPa are listed in Table~\ref{tab:refinement}. The refinements at other pressures were similar and can be found in the cif-file provided as Supplemental Material. We note that the ambient-pressure experiment was performed before loading the cell with the He gas. This may be the reason for the abrupt change in the Yb displacement parameters $U_{11}$ and $U_{33}$ between 0 and 0.78\,GPa (Fig.~\ref{fig:local}d).

\section{Computational results}
\label{app:dft}

\subsection{General methodology}
Non-magnetic DFT calculations were performed in the \texttt{FPLO}~\cite{fplo} and \texttt{VASP}~\cite{vasp1,*vasp2} codes. Atomic positions were optimized until the energy minimum was reached, and residual forces dropped below 0.005\,eV/\r A. The typical $k$-mesh included 64 points within the first Brillouin zone. Experimental lattice parameters were used at each pressure, and only atomic positions were relaxed.

\begin{table}
\caption{\label{tab:refinement}
Details of the single-crystal XRD data collection and structure refinement at 10\,GPa.
}
\begin{ruledtabular}
\begin{tabular}{cc}
 $T$ (K) & 298 \\
 $a$ (\r A) & 3.34272(13) \\
 $c$ (\r A) & 24.6033(7) \\
 wavelength (\r A) & 0.41114 \\
 $h_{\min}$, $h_{\max}$ & $-4\leq h\leq 3$ \\
 $k_{\min}$, $k_{\max}$ & $-5\leq k\leq 5$ \\
 $l_{\min}$, $l_{\max}$ & $-39\leq l\leq 38$ \\
 No. of reflections & 126 \\
 No. of refinable parameters & 9 \\
 $R_{\rm int}$ & 0.0249 \\
 $R_I$ & 0.0310 \\
\end{tabular}
\end{ruledtabular}
\end{table}

The results of structure optimization may be affected by the computational methodology. In Table~\ref{tab:geometry}, we analyze the role of three factors: i) band-structure code; ii) exchange-correlation potential; and iii) treatment of the Yb $4f$ shell. The calculations are performed for the simplest $R3m$ ordered model of YbMgGaO$_4$ described in App.~\ref{app:models} below.

The choice of the band-structure code affects the results of the relaxation. \texttt{VASP} calculations predict the too long Yb--O distances and, consequently, the too low angles. On the other hand, the relaxations within \texttt{FPLO} produce, irrespective of the methodology, the average distance $(d_1+d_2)/2$ and the average angle $(\beta_1+\beta_2)/2$ within, respectively, 0.01\,\r A and $0.5^{\circ}$ from the experimental values. The larger deviations of the \texttt{VASP} results are probably related to the lower accuracy of the default pseudopotential for the Yb atoms, whereas \texttt{FPLO} does not rely on pseudopotentials and introduces no approximations to the crystal potential.

\begin{table*}
\caption{\label{tab:geometry}
Comparison of the relaxed geometrical parameters for the ambient-pressure crystal structure in the primitive cell with the $R3m$ symmetry. Different band-structure codes (\texttt{FPLO} vs. \texttt{VASP}), exchange-correlation potentials (LDA~\cite{pw92} vs. GGA~\cite{pbe96}), and approaches to the treatment of the Yb $4f$ shell (valence vs. core) have been used.
}
\begin{ruledtabular}
\begin{tabular}{ccccccc}
 Code & \texttt{FPLO} & \texttt{FPLO} & \texttt{FPLO} & \texttt{FPLO} & \texttt{VASP} & Experiment \\
 $V_{\rm xc}$ & LDA   & LDA           & GGA           & GGA           & GGA           & \\\medskip
 Yb $4f$      & valence & core        & valence       & core          & core          & \\
 $d_1/d_2$ (\r A) & 2.229/2.248 & 2.214/2.233 & 2.231/2.257 & 2.215/2.243 & 2.258/2.270 & 2.234(4) \\
 $\beta_1/\beta_2$ (deg) & 98.62/99.73 & 99.52/100.68 & 98.09/99.60 & 98.88/100.62 & 97.31/98.02 & 99.44(13) \\
\end{tabular}
\end{ruledtabular}
\end{table*}

All choices mentioned here, including the \texttt{VASP} calculations, produce qualitatively similar results as a function of pressure. The trends obtained from DFT are thus robust. The results shown in the manuscript are obtained with \texttt{FPLO}, GGA functional, and Yb $4f$ states placed into the core, as this choice improved the convergence for larger unit cells. 

\subsection{Ordered structural models}
\label{app:models}
As DFT can not treat mixed sites, ordered structural models are used for calculations. We adopted three types of such ordered models:

1. $R3m$ model obtained from the parent $R\bar 3m$ structure by splitting the mixed Mg/Ga site into two, one fully occupied by Ga and the other one fully occupied by Mg. In this case, each Yb layer is sandwiched between the Mg and Ga layers leading to the off-center Yb displacement, two non-equivalent Yb--O distances $d_1\neq d_2$, and two O--Yb--O angles $\beta_1\neq\beta_2$. The $R3m$ structure was used in Table~\ref{tab:geometry} for testing the effect of band-structure code, basis set, and exchange-correlation potential.

2. $P3m$ models that preserve three-fold symmetry and feature layer-by-layer ordering of Mg and Ga. We chose two of such ordered structures (Fig.~\ref{fig:deformation}) that correspond to the third and fourth structures from Fig.~S5 of Ref.~\onlinecite{li2017a}.

In the structure from Fig.~\ref{fig:deformation}a, all Yb atoms develop a nearly symmetric local environment. This structure was used to evaluate pressure dependence of the Yb--O distances and Yb--O--Yb angles shown in Fig.~\ref{fig:structure}. In this case, the octahedra are still deformed, but $d_2-d_1<0.005$\,\r A and $\beta_2-\beta_1<0.2^{\circ}$ indicate a negligibly small deformation. The ``symmetric'' structure (left panel of Fig.~\ref{fig:deformation}) is also used to obtain the displacements ($\Delta z_{\rm Yb}$) of Yb within the octahedra B (GaMgYbMgGa). 

The structure from Fig.~\ref{fig:deformation}b covers the opposite scenario of the Yb atoms with a highly asymmetric local environment. It contains the octahedra A (GaGaYbMgMg) and C (GaGaYbMgGa) showing the largest values of $d_2-d_1$ and $\beta_2-\beta_1$. As explained in the main text, thermal displacement parameter of the average structure, $U_{33}$, does not reflect deformations of individual octahedra and shows instead relative displacements of the Yb atoms with respect to each other. In the ideal structure, the Yb atoms should be at $z=0,\frac13,\frac23$. To assess the effect of pressure on $U_{33}$, we compare the positions of the Yb atoms within the octahedra A and C to the remaining Yb atom at $z\simeq\frac23$ that shows a rather symmetric local environment and can be used as reference. The values of $\Delta z_{\rm Yb}$ are obtained as $\left|z_{A,C}-z_{\rm ref}-\frac13\right|$, where $z_{\rm ref}$ is the $z$-coordinate of the reference Yb atom.

3. $P3m$ model with the four-fold $2\mathbf a\times 2\mathbf b\times \mathbf c$ supercell, where $\mathbf a$, $\mathbf b$, and $\mathbf c$ are lattice vectors of the parent $R\bar 3m$ structure. This supercell allows to construct configurations, where adjacent Yb atoms have different local environment. We focus on the fragment shown in Fig.~\ref{fig:deformation}c, where Yb1--Yb2 and Yb2--Yb3 represent two non-equivalent superexchange pathways caused by the uneven distribution of Mg and Ga around the Yb layer.

\section{$\mu$SR experiment}
\label{app:musr}

\begin{table}
\caption{\label{tab:muSR}
Details of the $\mu$SR experiments.
}
\begin{ruledtabular}
\begin{tabular}{ccccc}\smallskip
 $P$ (GPa) & Spectrometer  & Pressure cell & $T$ (K) & Mode \\
 0   & Dolly & none        & 0.26--200 & ZF/LF \\
 0   & GPD   & MP35N+MP35N & 0.25--10  & ZF    \\
 1.9 & GPD   & MP35N+CuBe  & 0.25--10  & ZF    \\
 2.6 & GPD   & MP35N+MP35N & 0.25--10  & ZF/LF \\
\end{tabular}
\end{ruledtabular}
\end{table}

$\mu$SR measurements were performed in double-walled MP35 pressure cells with Daphne oil 7373 as pressure-transmitting medium. The pressure value was determined by measuring the superconducting transition of a small piece of indium positioned next to the sample inside the pressure cell. Further experimental details are summarized in Table~\ref{tab:muSR}

Two different pressure cells were used. The 0\,GPa and 2.6\,GPa data were collected in the double-walled MP35N+MP35N type cell~\cite{khasanov2016}, whereas the 1.9\,GPa data were collected separately in the low-background double-walled MP35N+CuBe type cell~\cite{shermadini2017}. For both cells, we employed Eq.~\eqref{eq:asym} and expressed the pressure-cell contribution as the Kubo-Toyabe depolarization function multiplied by an exponential damping,
\begin{equation*}
 A_{\rm PC}(t) = \left[\frac13+\frac23(1-\sigma_{\rm PC}^2t^2)e^{-\sigma_{\rm PC}^2t^2/2}\right]\,\exp(-\lambda_{\rm PC}t).
\end{equation*}

In the case of the MP35N+MP35N type cell, $\sigma_{\rm PC}=0.29\,\mu\text{s}^{-1}$ was temperature-independent, whereas $\lambda_{\rm PC}$ remained constant down to 1\,K and increased at lower temperatures, similar to Ref.~\onlinecite{khasanov2016}. The fraction of the signal coming from the pressure cell was $f_{\rm PC}=0.7$. In the case of the MP35N+CuBe type cell, both $\sigma_{\rm PC}=0.32\,\mu\text{s}^{-1}$ and $\lambda_{\rm PC}=0.025\,\mu\text{s}^{-1}$ are temperature-independent in agreement with those of Ref.~\onlinecite{shermadini2017}, and $f_{\rm PC}=0.5$. The usage of different cells may be the reason for a slight offset between the 1.9\,GPa data and 2.6\,GPa data in Fig.~\ref{fig:asymmetry}.

\begin{figure*}
\includegraphics[scale=0.9]{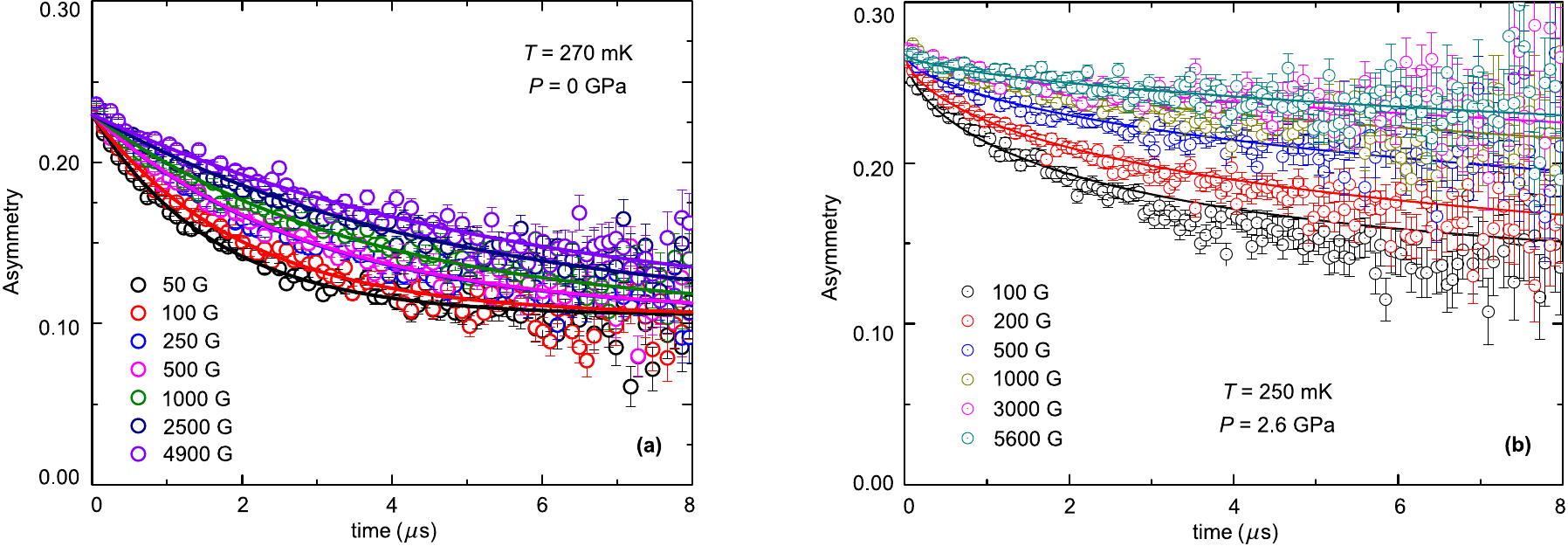}
\caption{\label{fig:amb}
Decoupling experiments (asymmetry as a function of time in different applied longitudinal fields) at ambient pressure and 270\,mK (a), and at 2.6\,GPa and 250\,mK (b).
}
\end{figure*}

In longitudinal fields above 100\,G, the contribution of the pressure cell should be mostly decoupled, and the data include only a constant, field-independent background. Fig.~\ref{fig:amb} shows the LF data fitted by the stretched exponentials to obtain $\lambda(H)$ presented in Fig.~\ref{fig:scaling}.

%

\end{document}